\documentclass[conference]{IEEEtran}
\IEEEoverridecommandlockouts
\usepackage{cite}
\usepackage{amsmath,amssymb,amsfonts}
\usepackage{algorithmic}
\usepackage{graphicx}
\usepackage{textcomp}
\usepackage{xcolor}

\definecolor{BestGreen}{RGB}{0,120,0}
\definecolor{SecondBlue}{RGB}{0,0,180}
\definecolor{WorstRed}{RGB}{180,0,0}

\def\BibTeX{{\rm B\kern-.05em{\sc i\kern-.025em b}\kern-.08em
    T\kern-.1667em\lower.7ex\hbox{E}\kern-.125emX}}
\begin{document}

\title{SPARX: Secure and Privacy-Aware Approximate CNN Acceleration with Edge RISC-V SoC}

\author{
\IEEEauthorblockN{Sonu Kumar}
\IEEEauthorblockA{\textit{Centre for Advanced Electronics} \\
\textit{Indian Institute of Technology Indore} \\
Simrol 453552, Indore, India \\
phd2101191002@iiti.ac.in}

\and 
 
\IEEEauthorblockN{Akash Sankhe, Mukul Lokhande, Santosh Kumar Vishvakarma}
\IEEEauthorblockA{\textit{Dept. of Electrical Engineering} \\
\textit{Indian Institute of Technology Indore} \\
Simrol 453552, Indore, India \\
\{ms2304102003, phd2201102020, skvishvakarma\}@iiti.ac.in}\\
\thanks{This work was supported partially by the Dept of Science and Technology (DST), Govt of India, for the INSPIRE PhD fellowship, and MeitY/SMDP-C2S for ASIC design tools.}
}

\maketitle

\begin{abstract}
Edge-AI systems increasingly require real-time CNN inference under strict energy, performance, security, and privacy constraints. Approximate computing improves hardware efficiency by exploiting the error resilience of neural network workloads; however, most approximate CNN accelerators do not jointly consider secure, privacy-aware edge deployment. This paper presents SPARX, a Secure and Privacy-Aware Approximate CNN Acceleration framework integrated within a heterogeneous RV32IMC RISC-V System-on-Chip (SoC). SPARX combines a custom RISC-V instruction extension, an approximate logarithmic CNN acceleration unit, a lightweight differential-noise-based privacy engine, and a challenge-response authentication mechanism. To guide arithmetic selection, an approximation-aware decision framework is introduced that uses the Approximation Severity Index (ASI), Approximation Efficiency (AE), Quality of Approximation (QoA), Approximation Figure-of-Merit (AFOM), and Hardware Acceleration Efficiency (HAE). Evaluation across 11 state-of-the-art approximate MAC architectures identifies the Iterative Logarithmic Multiplier (ILM) as the most suitable design, achieving 51.7\% area reduction, 81.5\% power reduction, and 2.13$\times$ throughput improvement compared with an accurate radix-4 Booth MAC, while only reducing ResNet-20/CIFAR-10 accuracy by 2.82 percentages. FPGA implementation on a Xilinx VC707 platform achieves 58.4 GOPS/W energy efficiency at 250 MHz, while 28-nm CMOS physical implementation validates ASIC feasibility.
\end{abstract}


\section{Introduction}

Edge intelligence enables real-time decision-making across resource-constrained platforms, including autonomous systems, intelligent surveillance, healthcare devices, augmented and virtual reality systems, and Internet of Things (IoT) nodes. These applications increasingly rely on Convolutional Neural Networks (CNNs) for perception and inference tasks, as demonstrated by recent advances in object detection, adverse-weather perception, medical image analysis, and visual analytics \cite{jin2025research, cao2025enhanced, hossain2023region, 11516214, wang2022novel}. However, modern CNN workloads impose significant computational and memory demands, making energy-efficient and low-latency hardware acceleration essential for edge deployment.

Specialised neural-network accelerators, vector-systolic architectures, and heterogeneous processor-accelerator platforms have been widely explored for efficient inference \cite{11014431, aguirre2024hardware, holla2026limo, VSA2025}. In this context, open and extensible RISC-V architectures provide a flexible platform for integrating domain-specific accelerators while preserving programmability, scalability, and deployment flexibility \cite{FlexPE2025, LPRE2025, XRNPE2026}. Consequently, heterogeneous RISC-V SoCs have emerged as attractive candidates for next-generation edge-AI systems.

\begin{figure}[!t]
\centering
\includegraphics[width=\columnwidth]{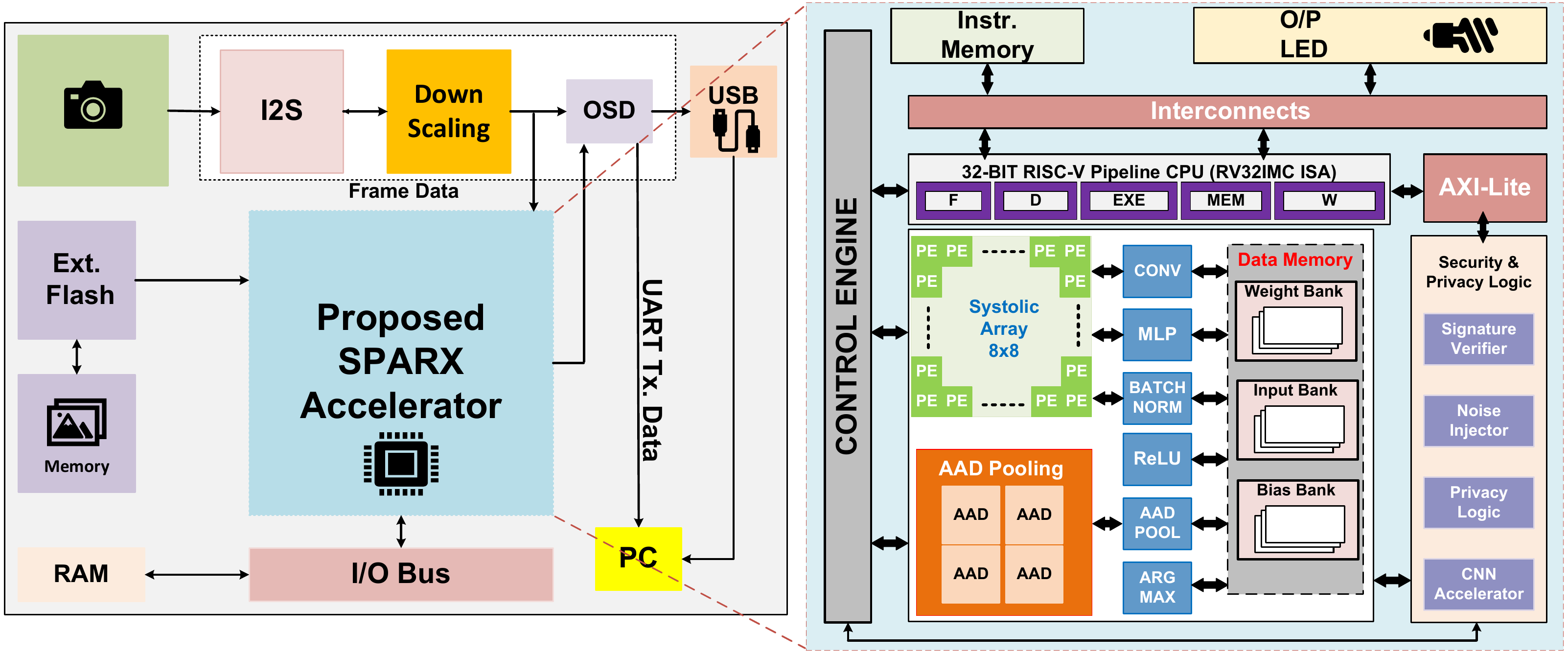}
\caption{Architecture of the proposed Secure and Privacy-Aware Approximate CNN Acceleration-enabled Edge RISC-V SoC on Artix-7 FPGA.}
\label{fig:secure_approx_soc}
\end{figure}

Approximate computing aims to reduce the cost of CNN accelerators by trading off strict arithmetic accuracy requirements. Since CNN workloads exhibit inherent error resilience, approximate arithmetic can reduce area, power, and latency while maintaining acceptable application-level accuracy \cite{ApproxComputingSurvey2025, ApproxMultiplierSurvey2024}. Prior work has explored approximate Booth multipliers, hybrid radix multipliers, dynamic-range-aware arithmetic, compressor-based designs, and logarithmic multiplication techniques for efficient AI inference \cite{R4ABM, RAD1024, DRUM, ROBA, LOBO, HRALM, MITCHELL_TRUNC, TL}. Furthermore, approximation-aware inference studies have shown that CNN layers exhibit different sensitivities to arithmetic errors, enabling hardware-software trade-offs between energy efficiency and accuracy \cite{HPRMul2024, HAMNN2025, sayadi2025layer, MARLIN2024}.

Despite these advances, most approximate CNN accelerators primarily focus on area, power, latency, and throughput \cite{kumar2026spade, kumar2026carmen, kumar2026corvet}, while security and privacy are often treated as secondary concerns. Edge devices routinely process sensitive visual, biometric, and contextual information, making them vulnerable to unauthorised access to accelerators, model misuse, memory snooping, and information leakage. Recent work \cite{sparsh} demonstrated lightweight privacy-preserving and authentication mechanisms within an RISC-V CNN accelerator. However, the interaction between approximation-aware arithmetic, privacy-aware execution, authentication, and accelerator-level performance remains largely unexplored.

To address this gap, this work proposes SPARX: Secure and Privacy-Aware Approximate CNN Acceleration with Edge RISC-V SoC. As illustrated in Fig.~\ref{fig:secure_approx_soc}, SPARX integrates an approximate logarithmic CNN accelerator, custom RISC-V instructions, differential-noise-based privacy protection, and challenge-response authentication within a heterogeneous RV32IMC RISC-V platform. The architecture enables runtime selection between exact, approximate, secure, and secure-approximate inference modes. In addition, an approximation-aware evaluation framework is introduced to systematically quantify arithmetic error, hardware efficiency, and accelerator-level performance.

The main contributions of this work are as follows:

\begin{itemize}

\item We propose SPARX (Secure and Privacy-Aware Approximate CNN Acceleration with Edge RISC-V SoC), a heterogeneous RV32IMC RISC-V platform that enables secure, privacy-aware, and energy-efficient CNN inference through runtime-selectable approximation and custom accelerator integration.

\item Custom RISC-V instruction extension and logarithmic MAC-based CNN accelerator are developed to support exact, approximate, secure, and secure-approximate inference modes. Compared with an accurate radix-4 Booth MAC, the selected arithmetic engine achieves 51.7\% area reduction, 81.5\% power reduction, and 2.13$\times$ throughput improvement, with only a 2.82 \% reduction in ResNet-20/CIFAR-10 inference accuracy.

\item Lightweight security framework that combines differential-noise-based privacy protection and challenge-response authentication is integrated into the accelerator datapath, enabling secure inference with minimal hardware cost. FPGA implementation on a Xilinx VC707 platform achieves 38.3 k-LUTs, 8.4 k-FFs, and 58.4 GOPS/W energy efficiency.

\item An approximation-aware evaluation framework based on ASI, AFOM, and HAE is introduced to systematically assess approximation-quality and hardware-efficiency trade-offs. Evaluation across 11 state-of-the-art approximate MAC architectures identifies ILM as the most suitable design, achieving an AFOM of 10.97 and an HAE of 2.56 while providing the best overall balance between accuracy, throughput, area, and power.

\end{itemize}

\section{SPARX System Architecture}

SPARX is a Secure and Privacy-Aware Approximate CNN Acceleration framework integrated within a heterogeneous RV32IMC RISC-V System-on-Chip (SoC). The architecture combines a tightly coupled CNN accelerator, runtime-selectable approximate arithmetic, lightweight privacy-preserving inference, and challenge-response authentication within a unified hardware-software co-design framework. Unlike conventional accelerator-coprocessor organisations, SPARX integrates CNN execution directly into the processor pipeline via custom instructions, enabling low-latency, energy-efficient edge-AI inference.

\begin{figure}[!t]
\centering
\includegraphics[width=\columnwidth]{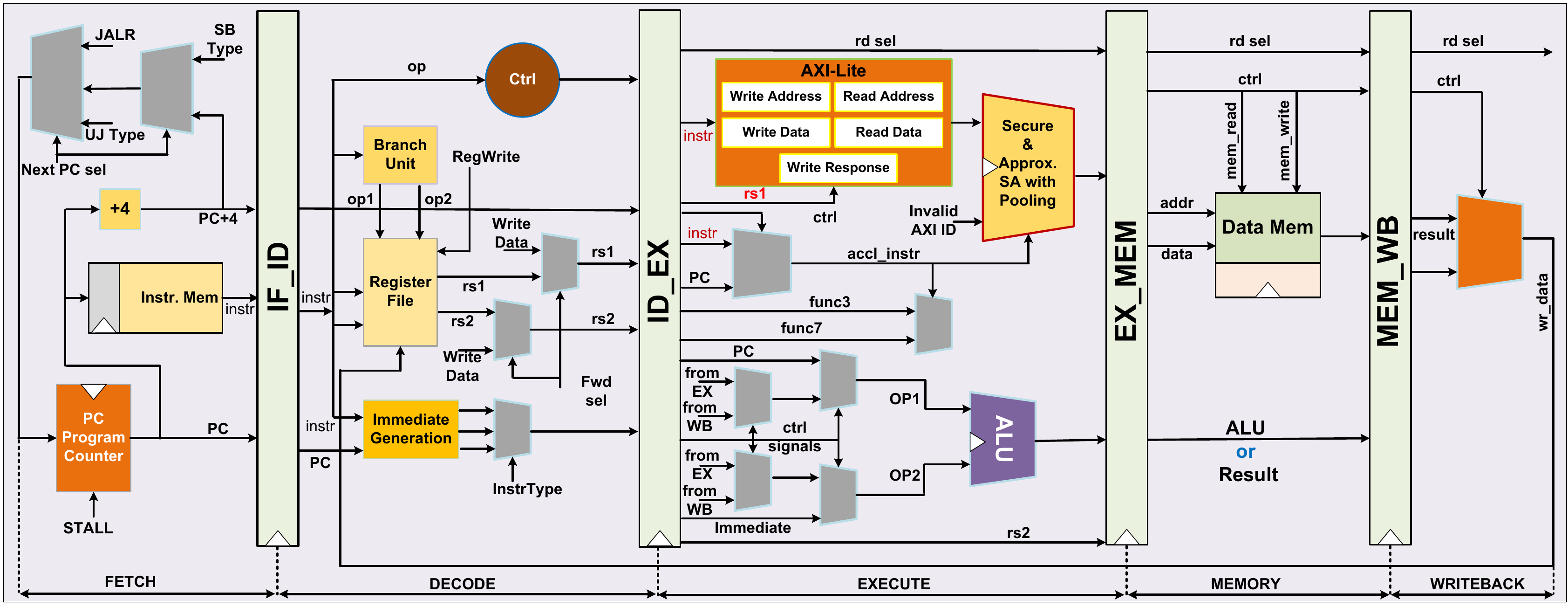}
\caption{Integration of the proposed Secure and Privacy-Aware Approximate CNN Acceleration into the RISC-V Execute (EX) stage, with custom M-extension for energy-efficient edge-AI inference.}
\label{fig:datapath}
\end{figure}

\begin{figure}[!t]
\centering
\includegraphics[width=\columnwidth]{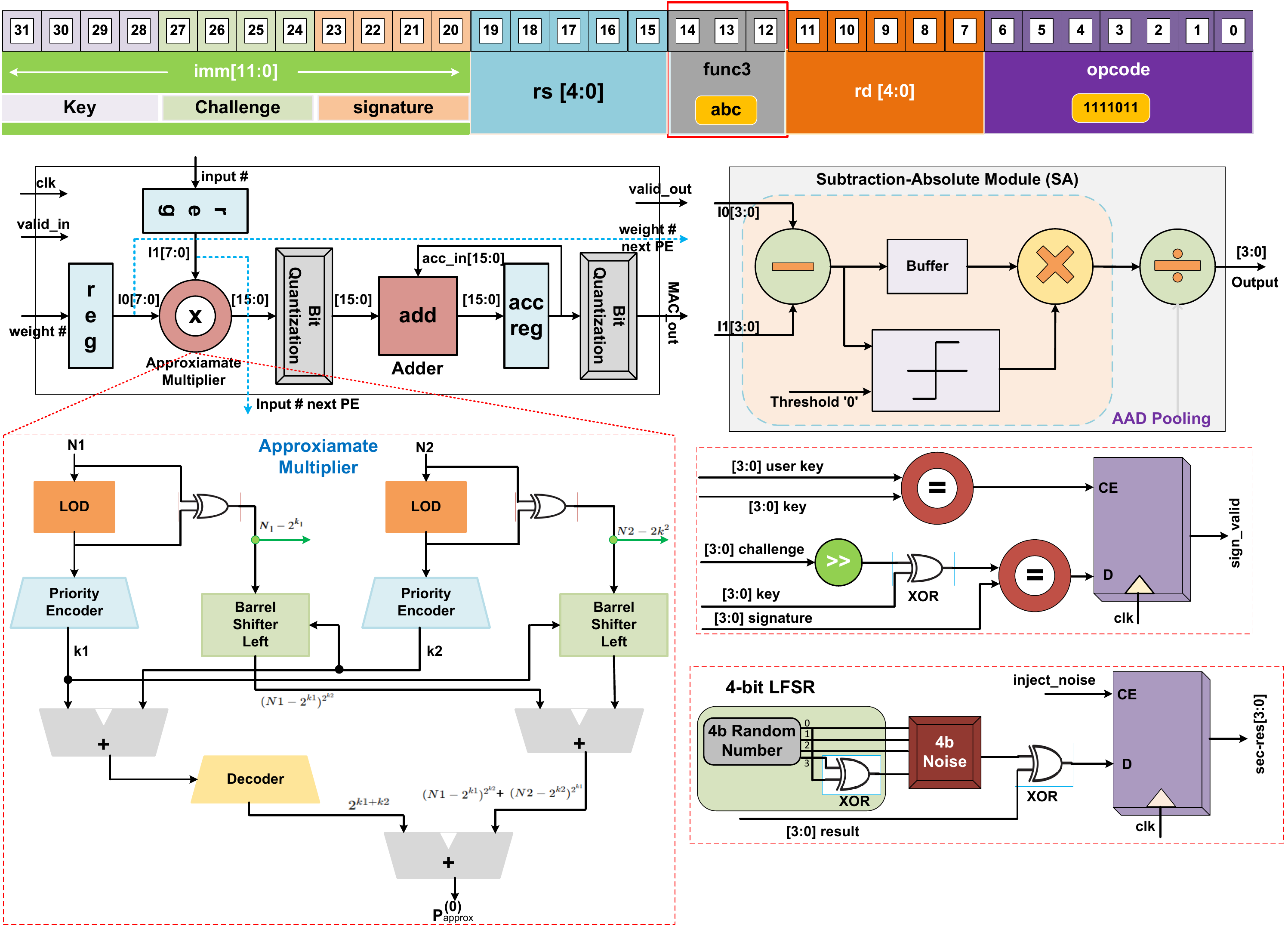}
\caption{(a) Custom RISC-V instruction encoding with an enhanced \texttt{func3} field (\texttt{abc}), where $a$ denotes privacy mode (0: disabled, 1: enabled), $b$ denotes approximation mode (0: exact, 1: approximate), and $c$ denotes the CNN variant (0: MNIST, 1: CIFAR-10). Accordingly, \texttt{000}/\texttt{001} correspond to MNIST/CIFAR-10, \texttt{100}/\texttt{101} to Secure MNIST/Secure CIFAR-10, \texttt{010}/\texttt{011} to Approximate MNIST/Approximate CIFAR-10, and \texttt{110}/\texttt{111} to Secure Approximate MNIST/Secure Approximate CIFAR-10, (b) Novel 8-bit approximate logarithmic multiplier and extended MAC unit, (c) Proposed 8-bit $2\times2$ AAD pooling, (d) Privacy-preserving inference enabled through differential noise injection, and (e) lightweight authentication module for signature-based verification during secure CNN inference.}
\label{fig:cnn_accelerator_building_blocks}
\end{figure}

The overall architecture is illustrated in Fig.~\ref{fig:secure_approx_soc}. The system consists of four primary components: (i) an RV32IMC RISC-V processor, (ii) an approximate CNN accelerator, (iii) a privacy and authentication engine, and (iv) an on-chip memory hierarchy. The RISC-V processor provides programmability and control, while the CNN accelerator performs computation-intensive inference operations. Dedicated BRAM memories store input feature maps, weights, biases, intermediate activations, and output feature maps to minimise external memory accesses and improve energy efficiency.

Figure~\ref{fig:datapath} illustrates the integration of the accelerator into the Execute (EX) stage of the processor pipeline. During instruction decoding, custom CNN instructions are identified and routed toward the accelerator interface. A dedicated multiplexer dynamically selects between the conventional ALU datapath and the accelerator datapath. Accelerator outputs are forwarded through the standard write-back path, preserving compatibility with the baseline RV32IMC architecture while significantly reducing software-managed accelerator invocation overhead.

\subsection{Custom Instruction Extension}

To support secure and approximation-aware execution, SPARX introduces the custom instruction format shown in Fig.~\ref{fig:cnn_accelerator_building_blocks}(a). A three-bit control field, denoted as \texttt{abc}, enables software-controlled selection of privacy mode, approximation mode, and CNN configuration.

The most significant bit ($a$) enables privacy-preserving execution. When asserted, accelerator outputs are routed through the differential-noise injection engine before being written back to the processor. The second bit ($b$) controls arithmetic precision. When $b=0$, the accelerator uses the exact MAC datapath, whereas $b=1$ activates the approximate logarithmic MAC datapath. The third bit ($c$) selects the CNN model configuration; $c=0$ corresponds to MNIST and $c=1$ to CIFAR-10 inference.

This encoding enables eight runtime-selectable operating modes spanning exact, approximate, secure, and secure-approximate inference without requiring hardware reconfiguration.

\subsection{Approximate CNN Acceleration Unit}

The computational core of SPARX consists of a CNN accelerator that supports convolution, activation, pooling, and fully connected operations required for 8-bit quantised CNN inference. Input feature maps are streamed from on-chip memory into dedicated MAC processing elements, while intermediate activations remain on-chip to maximise data reuse and reduce memory traffic.

The MAC processing element shown in Fig.~\ref{fig:cnn_accelerator_building_blocks}(b) operates in either exact or approximate mode depending on the instruction-controlled approximation bit. The approximate datapath employs the logarithmic multiplier shown in Fig.~\ref{fig:cnn_accelerator_building_blocks}(d), consisting of leading-one detectors, priority encoders, logarithmic converters, barrel shifters, and reconstruction logic. Multiplication is approximated in the logarithmic domain using addition and shift operations before accumulation.

Compared with conventional radix-4 Booth multiplication, the logarithmic architecture significantly reduces area, power consumption, and critical-path delay while maintaining acceptable CNN inference accuracy, thanks to the inherent error resilience of neural network workloads.

\subsection{Privacy and Authentication Engine}

To enhance inference security, SPARX incorporates lightweight privacy-preserving and authentication modules, as illustrated in Fig.~\ref{fig:cnn_accelerator_building_blocks}(e)-(f). The privacy engine employs a 4-bit Linear Feedback Shift Register (LFSR) to generate pseudo-random noise that is selectively injected into accelerator outputs. Let $Y_{cnn}$ denote the accelerator output and $N_{lfsr}$ represent the generated noise sequence. The privacy-preserving output is defined as

\begin{equation}
Y_{priv}=Y_{cnn}\oplus N_{lfsr}
\end{equation}

The injected perturbation obscures intermediate computational states, reduces information leakage through output observation or side-channel monitoring, and has a negligible impact on inference accuracy. In addition, a lightweight challenge-response authentication engine verifies execution requests before enabling accelerator operation. The authentication module receives a challenge, secret key, and signature, regenerates the expected signature, and grants execution access only when verification succeeds. This mechanism protects the accelerator from unauthorised access with minimal hardware overhead. With custom instruction extensions, approximate logarithmic arithmetic, privacy-preserving inference, and lightweight authentication within a unified accelerator framework, SPARX enables secure, flexible, and energy-efficient CNN inference for next-generation edge-AI systems.

\section{Unified Approximation-Aware Evaluation Framework}

SPARX extends the secure and privacy-aware RISC-V CNN accelerator paradigm\cite{sparsh} by introducing approximation-aware arithmetic and a unified methodology for selecting suitable approximate MAC architectures. Approximate computing improves hardware efficiency by exploiting the intrinsic error resilience of CNN workloads; however, approximation simultaneously affects multiple design objectives, including computational accuracy, silicon area, power consumption, throughput, and application-level inference quality. Consequently, selecting an appropriate arithmetic architecture requires evaluating both approximation quality and hardware efficiency.

Most prior studies evaluate approximate arithmetic using isolated metrics such as Normalised Mean Error Distance (NMED), Mean Absolute Error (MAE), Mean Squared Error (MSE), area, power, or delay. While useful individually, these metrics do not provide a unified assessment of approximation quality and implementation cost. To address this limitation, SPARX introduces an approximation-aware evaluation framework that combines arithmetic error characteristics, hardware implementation metrics, and accelerator-level performance indicators into a unified decision-making framework.

Let $Area_{base}$ and $Area_{approx}$ denote the ASIC area of the accurate radix-4 Booth MAC and the approximate design, respectively. Similarly, let $Power_{base}$ and $Power_{approx}$ represent the corresponding power consumption, while $Throughput_{base}$ and $Throughput_{approx}$ denote achievable throughput.

\subsection{Approximation Severity Index (ASI)}

To quantify approximation quality, the Approximation Severity Index (ASI) combines the Normalised Mean Error Distance (NMED), Mean Absolute Error (MAE), and Mean Squared Error (MSE) using a geometric mean:

\begin{equation}
ASI=
\sqrt[3]
{
\widehat{NMED}
\cdot
\widehat{MAE}
\cdot
\widehat{MSE}
}
\label{eq:asi}
\end{equation}

A lower ASI indicates lower approximation-induced degradation. The geometric mean ensures that poor performance in any individual error metric is appropriately penalised.

\subsection{Approximation Figure-of-Merit (AFOM)}

To jointly evaluate approximation quality and hardware efficiency, the Approximation Figure-of-Merit (AFOM) is defined as

\begin{equation}
AFOM=
\frac
{TOPS/W}
{
ASI
\cdot
\hat{Area}
}
\label{eq:afom}
\end{equation}

AFOM captures performance delivered per unit hardware cost and approximation error. Higher AFOM values indicate more efficient approximation strategies.

\subsection{Hardware Acceleration Efficiency (HAE)}

Hardware Acceleration Efficiency (HAE) jointly evaluates throughput improvement, hardware savings, and approximation quality.

\begin{equation}
TG=
\frac
{Throughput_{approx}}
{Throughput_{base}}
\end{equation}

\begin{equation}
AS = 1-\hat{Area},
\qquad
PS = 1-\hat{Power}
\end{equation}

\begin{equation}
HAE=
\frac
{
TG
\cdot
AS
\cdot
PS
}
{
ASI+\epsilon
}
\label{eq:hae}
\end{equation}

where $TG$, $AS$, and $PS$ represent throughput gain, area saving, and power saving, respectively. Higher HAE values indicate superior trade-offs between acceleration, hardware efficiency, and computational quality.

The proposed framework is subsequently used to evaluate multiple state-of-the-art approximate MAC architectures and identify the most suitable arithmetic engine for integration into the SPARX accelerator.

\section{Methodology and Hardware Evaluation}

This section evaluates representative approximate multiplication architectures using the proposed approximation-aware framework and validates the selected architecture through ASIC implementation, FPGA deployment, and CNN inference experiments. The objective is to identify the most suitable arithmetic engine for integration into the SPARX accelerator while balancing approximation quality, hardware efficiency, and application-level inference performance.

\begin{table}[!t]
\caption{Comparison for SoTA approximate and accurate MACs, ASIC metrics (area, power, and Freq.), ResNet-20/CIFAR-10 accuracy, and error metrics (NMED, MAE, and MSE).}
\label{tab:approx-mac}
\resizebox{\columnwidth}{!}{
\begin{tabular}{|l|c|c|c|c|c|c|c|}
\hline
\multicolumn{1}{|c|}{\textbf{Design}} & \textbf{Area ($\mu$m\textsuperscript{2})} & \textbf{Power (mW)} & \textbf{Freq. (MHz)} & Acc. (\%) & \textbf{NMED (10\textsuperscript{-3})} & \textbf{MAE (\%)} & \textbf{MSE (\%)} \\ \hline
\textbf{Accurate} & 526 & 58.43 & 147 & 87.23 & 0 & 0 & 0 \\ \hline
\textbf{HLR-BM\cite{HLR_BM}} & 406 & 40.03 & 178.6 & 85.3 & 17.8 & 7.2 & 3.66 \\ \hline
\textbf{AS-ROBA\cite{ROBA}} & 447 & 18.24 & 232.4 & 86.7 & 12.7 & 3.39 & 1.75 \\ \hline
\textbf{RAD1024\cite{RAD1024}} & 373 & 25.81 & 123.5 & 82.77 & 32.3 & 4.44 & 1.36 \\ \hline
\textbf{R4ABM\cite{R4ABM}} & 631 & 34.36 & 161 & 85.8 & 9.3 & 2.45 & 1.43 \\ \hline
\textbf{LOBO\cite{LOBO}} & 440 & 18.33 & 130 & 86.27 & 11.4 & 6.1 & 1.43 \\ \hline
\textbf{ROBA\cite{ROBA}} & 528 & 38.46 & 294 & 84.1 & 4.8 & 2.92 & 6.1 \\ \hline
\textbf{HRALM\cite{HRALM}} & 493 & 17.94 & 142.8 & 86.55 & 7.2 & 6.5 & 2.3 \\ \hline
\textbf{ALM-SOA\cite{ALM_SOA}} & 467 & 20.32 & 266 & 82.57 & 8.5 & 8.06 & 4.6 \\ \hline
\textbf{DRUM\cite{DRALM}} & 415 & 44.36 & 294 & 85.77 & 20.2 & 6.7 & 3.4 \\ \hline
\textbf{M-TRUNC\cite{MITCHELL_TRUNC}} & 387 & 19.31 & 221 & 85.12 & 23 & 14.43 & 1.47 \\ \hline
\textbf{ILM\cite{TL}} & 254 & 10.78 & 312.5 & 84.41 & 10.4 & 11.84 & 0.99 \\ \hline
\end{tabular}}
\end{table}

\subsection{Approximate MAC Selection}

To establish a representative design space, eleven state-of-the-art approximate multiplication architectures are implemented and evaluated alongside an accurate radix-4 Booth multiplier baseline. The evaluation considers arithmetic-error metrics (NMED, MAE, and MSE), hardware implementation metrics (area, power, and operating frequency), and application-level inference accuracy on ResNet-20/CIFAR-10.

\begin{table*}[!t]
\caption{Approximation-aware evaluation of representative approximate MAC architectures.}
\label{tab:result_metrics}
\resizebox{\textwidth}{!}{
\begin{tabular}{|l|r|r|r|r|r|r|r|r|r|r|r|r|}
\hline
\textbf{Design} & \textbf{AE\_A} & \textbf{AE\_P} & \textbf{QoA} & \textbf{ASI} & \textbf{Thrpt.} & \textbf{EE} & \textbf{EADPP} & \textbf{AFOM} & \textbf{TG} & \textbf{AS} & \textbf{PS} & \textbf{HAE} \\ \hline

\textbf{ILM} &
\textbf{\textcolor{BestGreen}{777.1325}} &
\textbf{\textcolor{BestGreen}{136.1410}} &
\textbf{\textcolor{BestGreen}{32.0697}} &
0.3500 &
\textbf{\textcolor{BestGreen}{20.0000}} &
\textbf{\textcolor{BestGreen}{1.8553}} &
\textbf{\textcolor{BestGreen}{3.0667}} &
\textbf{\textcolor{BestGreen}{10.9771}} &
\textbf{\textcolor{BestGreen}{2.1259}} &
\textbf{\textcolor{BestGreen}{0.5171}} &
\textbf{\textcolor{BestGreen}{0.8155}} &
\textbf{\textcolor{BestGreen}{2.5614}} \\ \hline

\textbf{AS-ROBA} &
264.9798 &
\textbf{\textcolor{SecondBlue}{134.8043}} &
\textbf{\textcolor{SecondBlue}{12.6437}} &
\textbf{\textcolor{SecondBlue}{0.2981}} &
14.8736 &
0.8154 &
\textbf{\textcolor{SecondBlue}{10.4582}} &
\textbf{\textcolor{SecondBlue}{3.2185}} &
1.5810 &
0.1502 &
0.6878 &
\textbf{\textcolor{SecondBlue}{0.5478}} \\ \hline

\textbf{MITCH\_TRUNC} &
250.1366 &
70.3981 &
7.4010 &
\textbf{\textcolor{WorstRed}{0.5557}} &
14.1440 &
0.7325 &
18.7906 &
1.7915 &
1.5034 &
0.2643 &
0.6695 &
0.4787 \\ \hline

\textbf{RAD1024} &
\textbf{\textcolor{SecondBlue}{373.7514}} &
79.6848 &
7.7986 &
0.4094 &
\textbf{\textcolor{WorstRed}{7.9040}} &
0.3062 &
31.9137 &
1.0549 &
\textbf{\textcolor{WorstRed}{0.8401}} &
\textbf{\textcolor{SecondBlue}{0.2909}} &
0.5583 &
0.3333 \\ \hline

\textbf{LOBO} &
262.9709 &
122.6178 &
11.6524 &
0.3270 &
8.3200 &
0.4539 &
20.2871 &
1.6592 &
0.8844 &
0.1635 &
0.6863 &
0.3034 \\ \hline

\textbf{ALM-SOA} &
122.8234 &
79.3356 &
6.7423 &
0.4804 &
17.0240 &
\textbf{\textcolor{SecondBlue}{0.8378}} &
17.1381 &
1.9644 &
1.8095 &
0.1122 &
0.6522 &
0.2756 \\ \hline

\textbf{DRUM} &
203.6827 &
\textbf{\textcolor{WorstRed}{25.8182}} &
\textbf{\textcolor{WorstRed}{3.0635}} &
0.5450 &
\textbf{\textcolor{SecondBlue}{18.8160}} &
0.4242 &
34.1263 &
0.9865 &
\textbf{\textcolor{SecondBlue}{2.0000}} &
0.2110 &
\textbf{\textcolor{WorstRed}{0.2408}} &
0.1865 \\ \hline

\textbf{HLR-BM} &
218.7944 &
33.5485 &
3.4480 &
0.5485 &
11.4304 &
\textbf{\textcolor{WorstRed}{0.2855}} &
\textbf{\textcolor{WorstRed}{49.9122}} &
\textbf{\textcolor{WorstRed}{0.6745}} &
1.2150 &
0.2281 &
0.3149 &
0.1591 \\ \hline

\textbf{HRALM} &
98.2778 &
120.5839 &
10.3489 &
0.3358 &
9.1392 &
0.5094 &
20.7980 &
1.6187 &
0.9714 &
0.0627 &
\textbf{\textcolor{SecondBlue}{0.6930}} &
0.1258 \\ \hline

\textbf{ROBA} &
-6.4315 &
64.2184 &
4.8670 &
0.3110 &
18.8160 &
0.4892 &
21.4811 &
1.5673 &
\textbf{\textcolor{SecondBlue}{2.0000}} &
-0.0038 &
0.3418 &
-0.0084 \\ \hline

\textbf{R4ABM} &
\textbf{\textcolor{WorstRed}{-465.7224}} &
106.7613 &
6.2875 &
\textbf{\textcolor{BestGreen}{0.2255}} &
10.3040 &
0.2999 &
30.3671 &
1.1088 &
1.0952 &
\textbf{\textcolor{WorstRed}{-0.1996}} &
0.4119 &
\textbf{\textcolor{WorstRed}{-0.3995}} \\ \hline

\end{tabular}}
\end{table*}

Table~\ref{tab:approx-mac} summarises the obtained results. The accurate radix-4 Booth MAC occupies 526~$\mu$m$^2$, consumes 58.43~mW, and operates at 147~MHz with 87.23\% inference accuracy. Among all approximate designs, ILM achieves the smallest silicon area (254~$\mu$m$^2$), corresponding to a 51.7\% reduction relative to the accurate implementation. ILM also provides the lowest power consumption (10.78~mW), resulting in an 81.5\% power reduction, while simultaneously achieving the highest operating frequency of 312.5~MHz, corresponding to a 2.13$\times$ throughput improvement.

Although certain architectures provide lower arithmetic error metrics, their hardware benefits are comparatively limited. For example, R4ABM exhibits low approximation error but requires a larger silicon area than the accurate baseline, whereas AS-ROBA preserves inference accuracy but achieves substantially lower hardware savings than ILM. Considering approximation quality, hardware efficiency, and accelerator-level performance jointly, ILM provides the most favourable overall trade-off and is therefore selected as the arithmetic core of the proposed SPARX accelerator. Accumulatively based on proposed approximation-aware metrics, particularly AFOM and HAE, the ILM architecture is selected as the arithmetic core for the proposed SPARX accelerator. ResNet-20/CIFAR-10 is used as a representative edge-vision benchmark to evaluate the proposed architecture.

\subsection{System-Level Implementation and Evaluation}

The selected ILM architecture was integrated into SPARX and validated through both FPGA and ASIC implementations. On a Xilinx VC707 FPGA, the approximate implementation reduces resource utilisation from 49.1k LUTs, 16.2k FFs, and 69 DSPs to 38.3k LUTs, 8.4k FFs, and 47 DSPs. Furthermore, the operating frequency increases from 62.78 MHz to 250 MHz (3.98$\times$), while energy efficiency improves from 10.3 GOPS/W to 58.4 GOPS/W (5.67$\times$), as summarised in Table~\ref{HW-arch}. Compared with representative FPGA-based CNN accelerators, SPARX achieves competitive hardware efficiency and demonstrates the effectiveness of approximation-aware arithmetic for edge-AI inference.

\begin{figure}[!t]
\centering
\includegraphics[width=0.35\columnwidth]{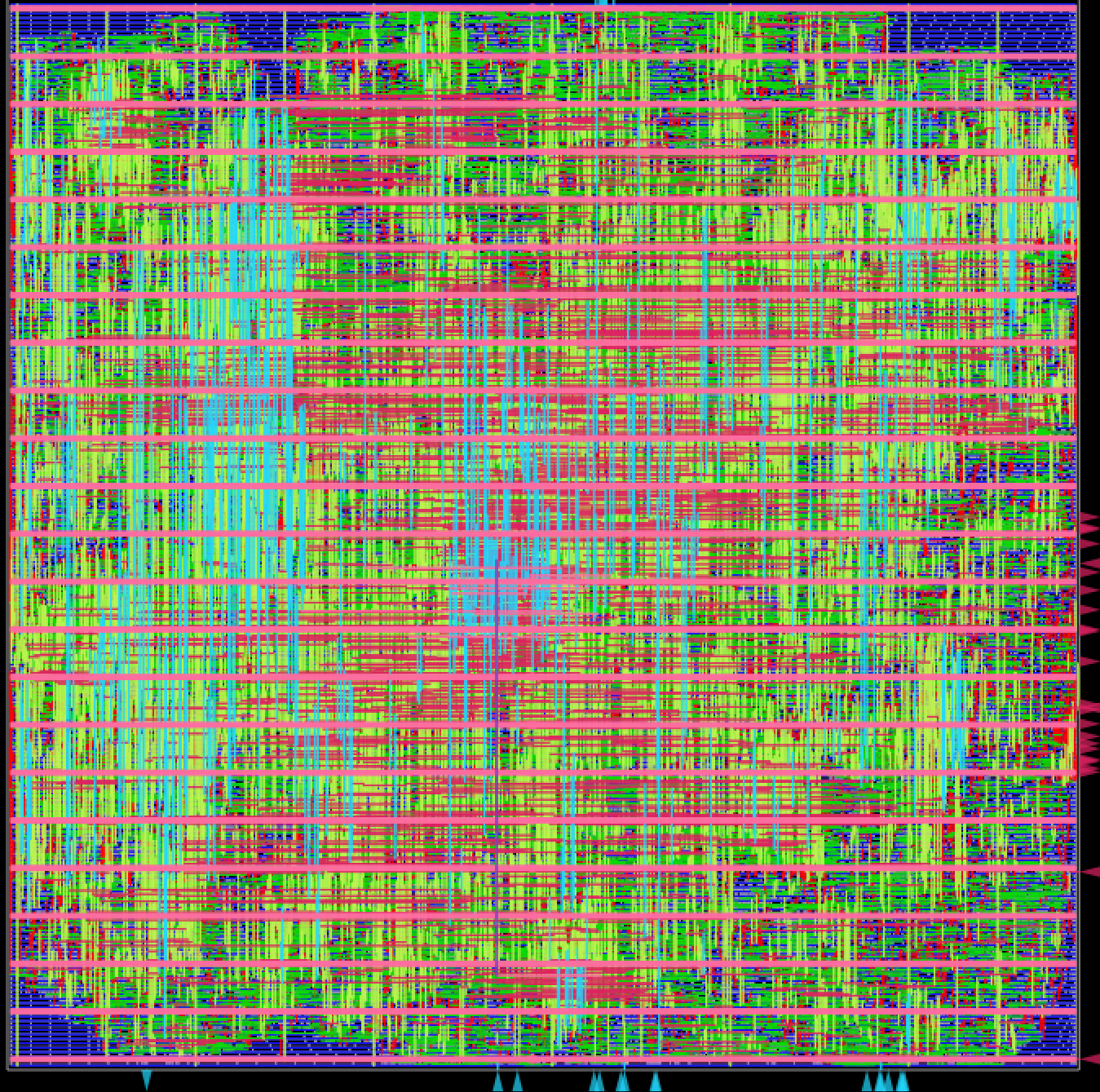}
\caption{28nm CMOS GDS view for the proposed SPARX-SoC.}
\label{fig:gds_layout_28nm}
\end{figure}

SPARX was synthesised and physically implemented in a 28-nm CMOS technology node. Successful post-layout implementation and GDS generation (Fig. \ref{fig:gds_layout_28nm}) confirm the practicality of integrating approximate logarithmic arithmetic, privacy-preserving logic, and lightweight authentication within a unified accelerator framework. These results demonstrate compatibility with standard ASIC design flows and validate the feasibility of secure and privacy-aware approximate acceleration for future edge-AI deployments.

\begin{table}[!t]
\caption{Comparison of FPGA-based CNN accelerator architectures.}
\label{HW-arch}
\resizebox{\columnwidth}{!}{%
\begin{tabular}{|l|cc|ccr|c|c|}
\hline
\textbf{Design} &
\textbf{Platform} &
\textbf{Model} &
\textbf{k-LUTs} &
\textbf{k-FFs} &
\textbf{DSPs} &
\textbf{\begin{tabular}[c]{@{}c@{}}Op. Freq\\(MHz)\end{tabular}} &
\textbf{\begin{tabular}[c]{@{}c@{}}Energy Efficiency\\(GOPS/W)\end{tabular}} \\
\hline

\textbf{TCAS-I'21~\cite{Xie-FlexAccl-TCASI21}}
& Arria10 & MobileNetV2
& 102 & 56 & 512 & 170 & 18.7 \\

\textbf{TCAS-I'22~\cite{DThanh-TCASI'22}}
& KCU15 & YoloV3-tiny
& 213 & 352 & 2240 & 200 & 10.3 \\

\textbf{TNNLS'19~\cite{NullHop-FlexAccl-TNNLS19}}
& Zynq7 & VGG16
& 229 & 107 & 128 & 60 & 27.5 \\

\textbf{TVLSI'25~\cite{FlexPE2025}}
& VC707 & Custom
& 38.7 & 7.4 & 73 & 466 & 8.42 \\

\textbf{Elsevier'2~\cite{TNPU}}
& VC707 & Custom
& 210 & 310 & 57 & 200 & 8 \\

\textbf{TCAS-I'24~\cite{Wu-Sp-systolic-TCASI24}}
& ZU3EG & ResNet-50
& 140.8 & 145.5 & 258 & 150 & 25 \\

\textbf{TCAS-II'23~\cite{SKi-TCASII'23}}
& XCVU9P & TinyYoloV3
& 132 & 39.5 & 96 & 150 & 6.36 \\

\textbf{HYDRA-1~\cite{HYDRA}}
& VC707 & Custom
& 115 & 115 & 32 & 100 & 4.5 \\

\textbf{TVLSI'20~\cite{Zhu-SparseCNN-TVLSI20}}
& ZCU102 & ResNet-50
& 390 & 278 & 1352 & 200 & 11.65 \\

\textbf{ESL'24~\cite{Yin-StrucSparse-ESL24}}
& ZCU102 & MobileNetV2
& 195 & 95.7 & 884 & 190 & 11.83 \\

\textbf{TVLSI'23~\cite{VaPr-WLee-TVLSI'23}}
& ZCU102 & XoR-Net
& 117 & 74 & 132 & 300 & 15.84 \\

\textbf{TCAD'23~\cite{WJiang-TCAD'23}}
& ZCU102 & MobileNet-v2
& 164 & 135 & 1283 & 333 & 34.4 \\

\hline

\textbf{This work (Acc)}
& VC707 & ResNet-20
& 49.1 & 16.2 & 69 & 62.78 & 10.3 \\

\textbf{This work (HLR-BM)}
& VC707 & ResNet-20
& 37.8 & 10.3 & 89 & 125 & 28.9 \\

\textbf{This work (ILM)}
& VC707 & ResNet-20
& 38.3 & 8.4 & 47 & 250 & 58.4 \\

\hline
\end{tabular}%
}
\end{table}

\section{Conclusion}

This paper presented SPARX, a Secure and Privacy-Aware Approximate CNN Acceleration framework integrated within a heterogeneous RV32IMC RISC-V SoC for energy-efficient edge-AI inference. The proposed architecture combines custom RISC-V instruction extensions, approximate logarithmic CNN acceleration, differential-noise-based privacy preservation, and lightweight challenge-response authentication within a unified accelerator framework. Evaluation across multiple state-of-the-art approximate MAC architectures identified ILM as the most suitable arithmetic engine, achieving 51.7\% area reduction, 81.5\% power reduction, and 2.13$\times$ throughput improvement compared with an accurate radix-4 Booth MAC. FPGA implementation on a Xilinx VC707 platform achieved 58.4 GOPS/W energy efficiency at 250 MHz, while a successful 28-nm CMOS implementation validated ASIC feasibility. These results demonstrate that SPARX effectively combines security, privacy, and hardware efficiency for next-generation edge-AI systems.

\bibliographystyle{ieeetr}
\bibliography{bib}

@INPROCEEDINGS{sparsh,
  author={Tyagi, Priyanshu and Patel, Rhythm and Mittal, Sparsh and Singhal, Rekha},
  booktitle={2026 39th International Conference on VLSI Design \& 25th International Conference on Embedded Systems (VLSID)}, 
  title={An FPGA-Based Secure and Privacy-Aware RISC-V SoC with a CNN Accelerator for Edge AI}, 
  year={2026},
  volume={},
  number={},
  pages={7-12}}

@article{cao2025enhanced,
  title={Enhanced YOLOv8 for accurate and efficient floating object detection on water surfaces},
  author={Cao, YanPeng and Luo, HaoWen and Wang, MengDi and Wang, Yue and Yan, Hao},
  journal={Scientific Reports},
  volume={16},
  pages={2907},
  year={2025}
}

@article{holla2026limo,
  title={LIMO: Low-power  matrix-multiplication primitive for edge computing},
  author={Holla, Amod and Chatterjee, Sumedh and Sen, Sutanu and Mukherjee, Anushka and Garc{\'\i}a-Redondo, Fernando and Biswas, Dwaipayan and Iacopi, Francesca and Roy, Kaushik},
  journal={npj Unconventional Computing},
  volume={3},
  number={1},
  pages={10},
  year={2026},
  publisher={Nature Publishing Group UK London}
}

@article{jin2025research,
  title={Research on target detection for autonomous driving based neural networks},
  author={Jin, Miao and Wang, Xiaohong and Guo, Ce and Yang, Shufan},
  journal={Scientific Reports},
  volume={15},
  number={1},
  pages={13725},
  year={2025},
  publisher={Nature Publishing Group UK London}
}

@article{hossain2023region,
  title={Region of interest (ROI) selection using vision transformer for automatic analysis using whole slide images},
  author={Hossain, Md Shakhawat and Shahriar, Galib Muhammad and Syeed, MM Mahbubul and Uddin, Mohammad Faisal and Hasan, Mahady and Shivam, Shingla and Advani, Suresh},
  journal={Scientific reports},
  volume={13},
  number={1},
  pages={11314},
  year={2023},
  publisher={Nature Publishing Group UK London}
}

@article{aguirre2024hardware,
  title={Hardware implementation of artificial neural networks},
  author={Aguirre, Fernando and Sebastian, Abu and Le Gallo, Manuel and Song, Wenhao and Wang, Tong and Yang, J Joshua and Lu, Wei and Chang, Meng-Fan and Ielmini, Daniele and Yang, Yuchao and others},
  journal={Nature communications},
  volume={15},
  number={1},
  pages={1974},
  year={2024},
  publisher={Nature Publishing Group UK London}
}

@article{sayadi2025layer,
  title={Layer-specific approximate multipliers for energy--precision trade-offs in convolutional neural networks},
  author={Sayadi, Ladan and Moaiyeri, Mohammad Hossein and Timarchi, Somayeh},
  journal={Scientific Reports},
  volume={15},
  number={1},
  pages={39482},
  year={2025},
  publisher={Nature Publishing Group UK London}
}

@article{wang2022novel,
  title={A novel optimized tiny YOLOv3 algorithm for the identification of objects in the lawn environment},
  author={Wang, Xinyan and Lv, Feng and Li, Lei and Yi, Zhengyang and Jiang, Quan},
  journal={Scientific Reports},
  volume={12},
  number={1},
  pages={15124},
  year={2022},
  publisher={Nature Publishing Group UK London}
}

@ARTICLE{FlexPE2025,
  author={Lokhande, Mukul and Raut, Gopal and Vishvakarma, Santosh Kumar},
  journal={IEEE Trans. Very Large Scale Integr. (VLSI) Syst.},
  title={{Flex-PE: Flexible and SIMD Multiprecision Processing Element for AI Workloads}},
  year={2025},
  volume={33},
  number={6},
  pages={1610-1623}
}

@INPROCEEDINGS{LPRE2025,
  author={Kokane, Omkar and Lokhande, Mukul and Raut, Gopal and Teman, Adam and Vishvakarma, Santosh Kumar},
  booktitle={Proc. IEEE Int. Symp. Circuits Syst. (ISCAS)},
  title={{LPRE: Logarithmic Posit-Enabled Reconfigurable Edge-AI Engine}},
  year={2025},
  pages={1-5}
}

@INPROCEEDINGS{XRNPE2026,
  author={Chaudhari, Tejas and Dewangan, Tanushree and Lokhande, Mukul and Vishvakarma, Santosh Kumar},
  booktitle={Proc. 39th Int. Conf. VLSI Design (VLSID)},
  title={{XR-NPE: High-Throughput Mixed-Precision SIMD NPE for Extended Reality Perception Workloads}},
  year={2026},
  pages={37-42}
}

@ARTICLE{HPRMul2024,
  author={Vafaei, Jafar and Akbari, Omid},
  journal={IEEE Trans. VLSI Syst.},
  title={{HPR-Mul: An Area and Energy-Efficient High-Precision Redundancy Multiplier by Approximate Computing}},
  year={2024},
  volume={32},
  number={11},
  pages={2012-2022}
}

@ARTICLE{HAMNN2025,
  author={Shi, Weiwei and Cao, Xiaocong and Zou, Zhuoliang and others},
  journal={IEEE Trans. VLSI Syst.},
  title={{Hybrid Approximate Multipliers With Merits Balance for Digital Processing and Neural Networks}},
  year={2025},
  volume={33},
  number={10},
  pages={2795-2805}
}

@ARTICLE{VSA2025,
  author={Li, Kai and Huang, Mingqiang and Li, Ang and others},
  journal={IEEE J. Solid-State Circuits},
  title={{A 29.12-TOPS/W Vector Systolic Accelerator With NAS-Optimized DNNs in 28-nm CMOS}},
  year={2025},
  volume={60},
  number={10},
  pages={3790-3801}
}

@ARTICLE{MARLIN2024,
  author={Guella, Flavia and others},
  journal={IEEE Trans. Circuits Syst. I},
  title={{MARLIN: A Co-Design Methodology for Approximate Reconfigurable Inference of Neural Networks at the Edge}},
  year={2024},
  volume={71},
  number={5},
  pages={2105-2118}
}

@article{ApproxMultiplierSurvey2024,
  title={A survey on approximate multiplier designs for energy efficiency: From algorithms to circuits},
  author={Wu, Ying and Chen, Chuangtao and Xiao, Weihua and Wang, Xuan and Wen, Chenyi and Han, Jie and Yin, Xunzhao and Qian, Weikang and Zhuo, Cheng},
  journal={ACM Transactions on Design Automation of Electronic Systems},
  volume={29},
  number={1},
  pages={1--37},
  year={2024},
  publisher={ACM New York, NY}
}

@ARTICLE{11516214,
  author={Sharma, Vijay Pratap and Venkatpurwar, Shekhar and Lokhande, Mukul and Pilipović, Ratko and Vishvakarma, Santosh Kumar},
  journal={IEEE Transactions on Computer-Aided Design of Integrated Circuits and Systems}, 
  title={ULTRA-MACE: A Unified Low-bit Trans-precision Reconfigurable Multiply-Accumulate Compute Engine for Accelerated Computing}, 
  year={2026},
  volume={},
  number={},
  pages={1-1}}

@article{ApproxComputingSurvey2025,
  title={Approximate computing survey, part i: Terminology and software \& hardware approximation techniques},
  author={Leon, Vasileios and Hanif, Muhammad Abdullah and Armeniakos, Giorgos and Jiao, Xun and Shafique, Muhammad and Pekmestzi, Kiamal and Soudris, Dimitrios},
  journal={ACM Computing Surveys},
  volume={57},
  number={7},
  pages={1--36},
  year={2025},
  publisher={ACM New York, NY}
}

@ARTICLE{R4ABM,
  author={Liu, Weiqiang and Qian, Liangyu and Wang, Chenghua and Jiang, Honglan and Han, Jie and Lombardi, Fabrizio},
  journal={IEEE Transactions on Computers},
  title={{Design of Approximate Radix-4 Booth Multipliers for Error-Tolerant Computing}},
  year={2017},
  volume={66},
  number={8},
  pages={1435-1441}
}

@ARTICLE{RAD1024,
  author={Leon, Vasileios and Zervakis, Georgios and Soudris, Dimitrios and Pekmestzi, Kiamal},
  journal={IEEE Transactions on Very Large Scale Integration (VLSI) Systems},
  title={{Approximate Hybrid High Radix Encoding for Energy-Efficient Inexact Multipliers}},
  year={2022},
  volume={30},
  number={3},
  pages={421-430}
}

@ARTICLE{HLR_BM,
  author={Waris, Haroon and Wang, Chenghua and Liu, Weiqiang},
  journal={IEEE Transactions on Circuits and Systems II: Express Briefs},
  title={{Hybrid Low Radix Encoding-Based Approximate Booth Multipliers}},
  year={2023},
  volume={70},
  number={12},
  pages={3367-3371}
}

@ARTICLE{LOBO,
  author={Ansari, Mohsen and Rehman, Safdar and Shafique, Muhammad and Henkel, J{\"o}rg},
  journal={IEEE Transactions on Computers},
  title={{On the Design of Logarithmic Multiplier Using Radix-4 Booth Encoding}},
  year={2020},
  volume={69},
  number={4},
  pages={499-510}
}

@ARTICLE{HRALM,
  author={Ansari, Mohsen and Rehman, Safdar and Shafique, Muhammad and Henkel, J{\"o}rg},
  journal={IEEE Transactions on Circuits and Systems I: Regular Papers},
  title={{A Hybrid Radix-4 and Approximate Logarithmic Multiplier for Energy Efficient Image Processing}},
  year={2018},
  volume={65},
  number={10},
  pages={3040-3053}
}

@INPROCEEDINGS{DRUM,
  author={Hashemi, Soheil and Bahar, R. Iris and Reda, Sherief},
  booktitle={Proceedings of the IEEE/ACM International Conference on Computer-Aided Design (ICCAD)},
  title={{DRUM: A Dynamic Range Unbiased Multiplier for Approximate Applications}},
  year={2015},
  pages={418-425}
}

@ARTICLE{ROBA,
  author={Zendegani, Reza and Kamal, Mehdi and Bahadori, Milad and Afzali-Kusha, Ali and Pedram, Massoud},
  journal={IEEE Transactions on Very Large Scale Integration (VLSI) Systems},
  title={{RoBA Multiplier: A Rounding-Based Approximate Multiplier for High-Speed yet Energy-Efficient Digital Signal Processing}},
  year={2017},
  volume={25},
  number={2},
  pages={393-401}
}

@ARTICLE{ALM_SOA,
  author={Liu, Weiqiang and Xu, Jiahua and Wang, Danye and Wang, Chenghua and Montuschi, Paolo and Lombardi, Fabrizio},
  journal={IEEE Transactions on Circuits and Systems I: Regular Papers},
  title={{Design and Evaluation of Approximate Logarithmic Multipliers for Low Power Error-Tolerant Applications}},
  year={2018},
  volume={65},
  number={9},
  pages={2856-2868}
}

@ARTICLE{MITCHELL_TRUNC,
  author={Kim, Min Soo and Del Barrio, Alberto A. and Oliveira, Leonardo Tavares and Hermida, Roman and Bagherzadeh, Nader},
  journal={IEEE Transactions on Computers},
  title={{Efficient Mitchell's Approximate Log Multipliers for Convolutional Neural Networks}},
  year={2019},
  volume={68},
  number={5},
  pages={660-675}
}

@ARTICLE{DRALM,
  author={Yin, Peipei and Wang, Chenghua and Waris, Haroon and Liu, Weiqiang and Han, Yinhe and Lombardi, Fabrizio},
  journal={IEEE Transactions on Sustainable Computing},
  title={{Design and Analysis of Energy-Efficient Dynamic Range Approximate Logarithmic Multipliers for Machine Learning}},
  year={2021},
  volume={6},
  number={4},
  pages={612-625}
}

@ARTICLE{TL,
  author={Pilipovi\'{c}, Ratko and Buli\'{c}, Patricio and Lotri\v{c}, Uro\v{s}},
  journal={IEEE Transactions on Circuits and Systems I: Regular Papers},
  title={{A Two-Stage Operand Trimming Approximate Logarithmic Multiplier}},
  year={2021},
  volume={68},
  number={6},
  pages={2535-2545}
}

@INPROCEEDINGS{11014431,
  author={Sankhe, Akash and Lokhande, Mukul and Sharma, Radheshyam and Vishvakarma, Santosh Kumar},
  booktitle={2025 26th International Symposium on Quality Electronic Design (ISQED)}, 
  title={Area-Optimized 2D Interleaved Adder Tree Design for Sparse DCIM Edge Processing}, 
  year={2025},
  volume={},
  number={},
  pages={1-6}}

@ARTICLE{NullHop-FlexAccl-TNNLS19,
  author={Aimar, Alessandro and Mostafa, Hesham and Calabrese, Enrico and Rios-Navarro, Antonio and Tapiador-Morales, Ricardo and others},
  journal={IEEE Transactions on Neural Networks and Learning Systems}, 
  title={{NullHop: A Flexible Convolutional Neural Network Accelerator Based on Sparse Representations of Feature Maps}}, 
month=mar,
  year={2019},
  volume={30},
  number={3},
  pages={644-656}}

@ARTICLE{Zhu-SparseCNN-TVLSI20,
  author={Zhu, Chaoyang and Huang, Kejie and Yang, Shuyuan and Zhu, Ziqi and Zhang, Hejia and Shen, Haibin},
  journal={IEEE Transactions on Very Large Scale Integration (VLSI) Systems}, 
  title={{An Efficient Hardware Accelerator for Structured Sparse Convolutional Neural Networks on FPGAs}},
month=sep, 
  year={2020},
  volume={28},
  number={9},
  pages={1953-1965}}

@ARTICLE{Xie-FlexAccl-TCASI21,
  author={Xie, Xiaoru and Lin, Jun and Wang, Zhongfeng and Wei, Jinghe},
  journal={IEEE Transactions on Circuits and Systems I: Regular Papers}, 
  title={An Efficient and Flexible Accelerator Design for Sparse CNNs}, 
month = jul,
  year={2021},
  volume={68},
  number={7},
  pages={2936-2949}}

@ARTICLE{Yin-StrucSparse-ESL24,
  author={Yin, Xiaodi and Wu, Zhipeng and Li, Dejian and Shen, Chongfei and Liu, Yu},
  journal={IEEE Embedded Systems Letters}, 
  title={{An Efficient Hardware Accelerator for Block Sparse Convolutional Neural Networks on FPGA}}, 
  year={2024},
  volume={16},
  number={2},
  pages={158-161}}

@ARTICLE{Wu-Sp-systolic-TCASI24,
  author={Wu, Bi and Yu, Tianyang and Chen, Ke and Liu, Weiqiang},
  journal={IEEE Transactions on Circuits and Systems I: Regular Papers}, 
  title={{Edge-Side Fine-Grained Sparse CNN Accelerator With Efficient Dynamic Pruning Scheme}}, 
month= mar,
  year={2024},
  volume={71},
  number={3},
  pages={1285-1298}}

@INPROCEEDINGS{HYDRA,
  author={Kumar, Sonu and Gupta, Komal and Dasanayake, Isuru S and Lokhande, Mukul and Vishvakarma, Santosh Kumar},
  booktitle={2025 IEEE 19th International Conference on Industrial and Information Systems (ICIIS)}, 
  title={HYDRA: A Resource-Efficient Hybrid Data-Multiplexed, Run-time Layer-Reconfigurable Compute Engine for DNN Acceleration}, 
  year={2026},
  volume={19},
  number={},
  pages={212-217}}

@article{TNPU,
title = {Trans-precision NPU for resource-efficient mobile AI acceleration},
author = {Mukul Lokhande and Vijay P. Sharma and S.V. Jaya Chand and others},
journal = {Journal of Systems Architecture},
volume = {177},
pages = {103866},
year = {2026},
issn = {1383-7621}}

@article{kumar2026carmen,
  title={CARMEN: CORDIC-Accelerated Resource-Efficient Multi-Precision Inference Engine for Deep Learning},
  author={Kumar, Sonu and Lokhande, Mukul and Vishvakarma, Santosh Kumar and Teman, Adam},
  journal={arXiv preprint arXiv:2605.06878},
  year={2026}
}

@article{kumar2026corvet,
  title={CORVET: A CORDIC-Powered, Resource-Frugal Mixed-Precision Vector Processing Engine for High-Throughput AIoT applications},
  author={Kumar, Sonu and Khan, Mohd Faisal and Lokhande, Mukul and Vishvakarma, Santosh Kumar},
  journal={arXiv preprint arXiv:2602.19268},
  year={2026}
}

@article{kumar2026spade,
  title={SPADE: A SIMD Posit-enabled compute engine for Accelerating DNN Efficiency},
  author={Kumar, Sonu and Vinnakota, Lavanya and Lokhande, Mukul and Vishvakarma, Santosh Kumar and Teman, Adam},
  journal={arXiv preprint arXiv:2601.17279},
  year={2026}
}

\end{document}